\documentclass[pre,reprint]{revtex4-1}

\newcommand{\rmd}{\ensuremath{\mathrm{d}}}

\newcommand{\abs}[1]{\ensuremath{\left\vert#1\right\vert}}
\newcommand{\e}{\mathrm{e}}
\usepackage[lofdepth,lotdepth,caption=false]{subfig}
\usepackage{amsmath}
\usepackage{amssymb}
\usepackage{mathtools}
\usepackage{graphicx}
\usepackage{dcolumn}
\usepackage{color}
\usepackage{blindtext}

\begin{document}

\title{Infrared spectroscopy of surface charges in plasma-facing dielectrics}
\author{K. Rasek, F. X. Bronold and H. Fehske}
\date{\today}

\address{Institut f{\"u}r Physik,
	Universit{\"a}t Greifswald, 17489 Greifswald, Germany }

\begin{abstract}
	We propose to measure the surface charge accumulating at the interface between a plasma and a dielectric by infrared spectroscopy
	using the dielectric as a multi-internal reflection element. The surplus charge leads to an attenuation of the transmitted signal from which the magnitude of the charge can be inferred. Calculating the optical response perturbatively in first order from the Boltzmann equation for the electron-hole plasma inside the solid, we can show that in the parameter range of interest a classical Drude term results. Only the integrated surface charge enters, opening up  thereby a very efficient analysis of measured data.

\end{abstract}

\maketitle

\section{Introduction}

Whenever two materials with different electronic structures are in contact, an electric double layer forms. Electrons flow from the material with the higher Fermi level to the other material, causing a charge imbalance. The electric field, resulting from this inhomogeneous charge distribution, or, equivalently, the band bending,  slows the flux of electrons until a balance is reached. This fundamental mechanism occurs at any interface between materials with (quasi) free electrons. 

An interface that falls into this category is the solid-bound plasma. 
While many physical and chemical processes occur at the plasma-solid interface~\cite{BFA19}, in this work we focus on the electronic interaction between the solid and the plasma~\cite{BRF20}. A microscopic theory of the electronic kinetics at the interface is desirable, for example, for catalysis~\cite{NOS15}. 
An integral part of this analysis should be experimental tools to measure the charges accumulated in the wall. Some methods for this purpose already exist, using electric probes~\cite{KA80}, optomechanical sensors~\cite{PFS96}, or the optopelectric Pockels effect~\cite{SVB19}. We propose that measurement of the surface charges is also possible by infrared spectroscopy.

Our previous proposal for this utilizes a stack of thin layers on top of a prism in order to utilize the Berreman resonance in the plasma-facing material, for which this layer may not be too thick~\cite{RBB18}.
The drawback of such a configuration is the need of at least two additional layers of materials, a metal and a dielectric, between the plasma and the prism used for the optics. The current method overcomes this complexity by employing the plasma-facing dielectric itself as a prism in which multiple reflection at the same interface occur.

\begin{figure}[t]
	\centering
	\includegraphics[width=1\linewidth]{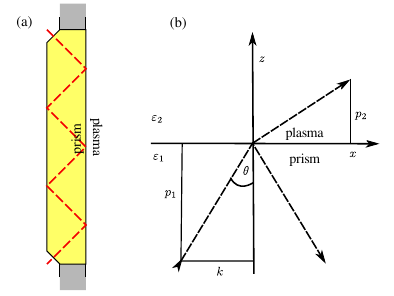}
	\caption{(a): Scheme of the multi-internal reflection element, here shown for $N=3$ internal reflections. The light (dashed line) is internally reflected $N$ times by the prism (yellow) at the plasma interface, and $N-1$ times at the opposite side, which is assumed to be perfectly reflecting. 
		(b): Geometry of the interface. The interface is located at $z=0$ as an abrupt change from material 1 ($z<0$, dielectric function $\varepsilon_1$) to material 2 ($z>0$, dielectric function $\varepsilon_2$). The incident wave vector has an angle $\theta$ from the $z$-axis and lateral (normal) component $k$ ($p_1$). The lateral component $k$ is preserved across the interface, and the normal component of the refracted wave is $p_2$. In case of total reflection $p_2$ is imaginary. 
	}
	\label{fig:overview}
\end{figure}

The setup, where incident IR light is reflected internally multiple times at the plasma-solid interface~\cite{NM97}, is shown in Fig.~\ref{fig:overview}. By measuring the transmitivity of such a multi-internal reflection element (MIRE) with and without the plasma, the accumulated wall charge within the prism material can be determined.

We have developed a microscopic kinetic theory for the plasma-solid interface~\cite{RBF20} based on the Boltzmann equation for the kinetic modeling of the charge carriers and the Poisson equation for the calculation of the potential and electric field. It presents a self-consistent model of electrons and ions impinging on the dielectric wall, where electrons are absorbed into the conduction band and ions create holes in the valence band. The kinetics of the charge carriers in the solid is governed by collision processes and the electric field, which in turn is determined by the density of the charge carriers. 
We have presented a method to self-consistently solve this set of equations and in the current work use that solution to calculate the optical response of the interface to infrared radiation. 

From the (change in) conductivity caused by the surplus charge deposited into the wall material by the plasma, we obtain a correction to the Fresnel reflectivity, using an approach developed by Feibelman~\cite{Feibelman82} and previously applied to metal surfaces~\cite{KG85}. 
It is based on two surface response functions $d_\perp$ and $d_\parallel$ in which all surface effects are incorporated, and which, in the long wavelength limit, are calculated by integrating over the change in the dielectric function. A similar theory had also been developed by Flores, Garcia-Moliner and Navascues to treat inhomogeneous interfaces and to incorporate surface boundary conditions into the standard Fresnel formalism~\cite{FGN71,FGN73}.  Although in general a nonlocal dielectric response defines the surface response functions, for the parameters of the plasma wall a local approximation is justified and leads to the Drude equation for the conductivity modifying the dielectric function. 

We suggested this setup as part of a general perspective on the electronic properties of the plasma-solid interface~\cite{BRF20}, demonstrating its viability however only by an exploratory calculation. In this work we examine the proposal in more detail. Solving the Boltzmann equation in the flat-band approximation, we moreover demonstrate that the distribution functions are effectively Maxwellian. By using the method of dominant balance for calculating the optical response, the conductivity becomes a Drude-like expression, allowing hence a straightforward analysis of experimental data. 

The paper is structured as follows. In Sec.~\ref{sec:Theory} we describe the theoretical framework of the dielectric response of the plasma-solid interface in calculating the transmitivity of the MIRE. Numerical results are presented in Sec.~\ref{sec:results} and Sec.~\ref{sec:conclusion} concludes the paper.

\section{Theoretical Framework}\label{sec:Theory}

In a MIRE as shown in Fig.~\ref{fig:overview}(a), the incident light is reflected several times at the plasma-solid interface of interest. The angle of incidence is chosen such that total internal reflection occurs at the interface, so that the reflection coefficient $R_s$ at this surface is close to one and high numbers of reflections $N$ are possible. With a finite reflectivity $R_0$ at the points of perpendicular entry and exit of the light, at the top and bottom of Fig.~\ref{fig:overview}(a), and with no coherence of the different reflexes, the transmitivity of the MIRE is~\cite{Milosevic86}
\begin{equation}
\label{eq:T}
T =\frac{R_s^N(1-R_0)^2}{1 - R_s^{2N}R_0^2}~.
\end{equation}

In the model we propose, the change in the optical response of the solid ($z<0$; index 1), caused by its interaction with the plasma ($z>0$; index 2) can be encapsulated in two surface response functions. They are determined by the deviation of the dielectric function near the interface from the bulk values $\varepsilon_{1,2}$~\cite{Feibelman82}. For p-polarized light, the two surface response functions, defined by
\begin{equation}\label{eq:d_para}
d_\parallel = \frac{1}{\varepsilon_1 - \varepsilon_2}\int_{-\infty}^{\infty}\rmd z\left\{ \varepsilon_{xx}(z) - \left[  \varepsilon_1\Theta(-z) + \varepsilon_2\Theta(z) \right]\right\}
\end{equation}
and 
\begin{equation}\label{eq:d_perp}
d_\perp = \frac{1}{\varepsilon_1^{-1} - \varepsilon_2^{-1}}\int_{-\infty}^{\infty}\rmd z\left\{ \varepsilon_{zz}^{-1}(z) - \left[  \frac{\Theta(-z)}{\varepsilon_1} + \frac{\Theta(z)}{\varepsilon_2} \right]\right\}~,
\end{equation}
modify the Fresnel reflectivity of the interface
\begin{equation}\label{eq:r}
r_0 = \frac{\varepsilon_1 p_2 - \varepsilon_2 p_1}{\varepsilon_1 p_2 + \varepsilon_2 p_1}
\end{equation}
through the correction factor 
\begin{equation}\label{eq:C}
C = 2 i p_1\frac{p_2^2\varepsilon_1d_\parallel - k^2\varepsilon_2d_\perp}{p_2^2\varepsilon_1 - k^2\varepsilon_2}~,
\end{equation}
according to
\begin{equation}
r = r_0(1+C)~,
\end{equation}
where $k = \sin{\theta} \sqrt{\varepsilon_1}\,\omega/c$, $p_1 = \cos{\theta}\sqrt{\varepsilon_1}\,\omega/c$, and $p_2 = \sqrt{\varepsilon_2 - \cos^2{\theta} \,\varepsilon_1 }\,\omega/c$, see also Fig.~\ref{fig:overview}(b). Consequently, the reflection coefficient of the interface $R_s = \abs{r}^2$.
In the long wavelength limit the dielectric functions in Eqs.~\eqref{eq:d_para} and~\eqref{eq:d_perp} can be approximated by~\cite{FG86}
\begin{align}
\label{eq:epsxx(z)}
\varepsilon_{xx}(z) &= \int_{-\infty}^{\infty}\rmd z' \varepsilon_{xx}(z,z')\\
\label{eq:epszz(z)}
\varepsilon^{-1}_{zz}(z) &= \int_{-\infty}^{\infty}\rmd z' \varepsilon_{zz}^{-1}(z,z')~.
\end{align}

Since in the long wavelength limit the nondiagonal elements of the dielectric tensor are neglegible, the inverse is given by
\begin{equation}
\label{eq:inv_eps}
\int_{-\infty}^{\infty}\rmd z'' \varepsilon_{zz}^{-1}(z,z'') \varepsilon_{zz}(z'',z') = \delta(z-z')~.
\end{equation}

The nonlocal dielectric tensor is
\begin{equation}\label{eq:eps_tens}
\underline{\underline{\varepsilon}}(z,z') = \varepsilon\delta(z-z') \underline{\underline{1}}+ i\frac{4\pi}{\omega}\underline{\underline{\sigma}}(z,z')~,
\end{equation} 
with the conductivity tensor defined by
\begin{equation}
\label{eq:j=simgaE}
{\bf j}(z, {\bf k}_F) = \int_{z}^0 \rmd z'\underline{\underline{\sigma}}(z,z',{\bf k}_F){\bf E}(z',{\bf k}_F)~,
\end{equation}
where ${\bf E}$ is the external electric field of the incident infrared light with wavevector ${\bf k}_F$, and ${\bf j}$ is the electric current.
Here and in the following $\varepsilon = \varepsilon_1$ is the bulk dielectric function of the wall material, $\varepsilon_2 = 1$, and the influence of the plasma on the optical response is neglected, besides providing the surplus charges. 
We imply homogeneity in the directions parallel to the interface, hence only $z$ appears in the equations as a spatial variable. 

The Boltzmann equation, describing the distribution function $F_s$ for electrons in the conduction band ($s=*$) and holes in the valence band ($s=h$), is
\begin{align}\label{eq:BE}
\left(\partial_t  + {{\bf v}}_s \cdot\nabla_{{\bf r}} +\dot{{\bf k}}\cdot  \nabla_{{\bf k}} + \gamma_s  \right)F_s = \Phi_s~,
\end{align}
with
\begin{equation}
\dot{\bf{k}} =  q_s(\mathcal{E}\hat{\bf z} + {\bf{E}})~,
\end{equation}
where $\hat{\bf z}$ is the unit vector in $z$ direction, ${\bf{v}}_s$ is the velocity, $q_s$ is the charge of the species ($q_h=1$ for holes, $q_*=-1$ for electrons) and $\mathcal{E}$ is the intrinsic electric field caused by the surplus charges and determined by the once integrated Poisson equation
\begin{equation}\label{eq:PE}
\mathcal{E}(z)  = \frac{16 \pi}{\varepsilon}\int_{-\infty}^{z}\rmd z 
\left[n_e(z)-n_h(z)\right] ~~.
\end{equation}
Collision integrals are described in Eq.~\eqref{eq:BE} by the terms $\Phi$, representing scattering into state $F_s({\bf{r}},{\bf{k}},t)$, and $\gamma_s$, describing scattering out of $F_s$. 

By finding the linear relation between the external electric field ${\bf E}$ and the flux ${\bf j}$ that it causes, the conductivity tensor $\underline{\underline{\sigma}}$ can be identified through Eq.~\eqref{eq:j=simgaE}. 
For that purpose we utilize the solution of the Boltzmann equation and Poisson equation for the plasma-solid interface that we previously reported~\cite{RBF20}. The external electric field due to the infrared light is viewed as a perturbation to that solution. To find the linear response, we expand the distribution function to first order in the perturbing field ${\bf E}$,
\begin{equation}
\label{eq:F_expansion}
	F_s(z) = F_0(z) + \int\rmd z' {\bf{F}}_1(z,z')\cdot {\bf{E}}(z') + \mathcal{O}({\bf E}^2)~.
\end{equation}
Similarly, $\mathcal{E}$ and $\Phi_s$ are expanded.
Then, the unperturbed, static solution is determined by
\begin{equation}\label{eq:BE_0}
	\left(v_z \partial_z + q \mathcal{E}_0 \partial_{k_z} +  \gamma_0  \right)F_0  = \Phi_{0}~,
\end{equation}
and for the linear response the Boltzmann equation yields
\begin{multline}\label{eq:BE_1}
	\left(\partial_t  + {\bf{v} }\cdot\nabla_{\bf{r}} + q \mathcal{E}_0 \partial_{k_z} +  \gamma_0  \right){\bf{F}}_1 \\
	= {\bf{\Phi}}_{1} - q\left[ {\bf{\mathcal{E}}}_1\cdot\hat{\bf{z}} + \delta(z-z')\right]\nabla_{\bf{k}}F_0 + \mathcal{O}({\bf{E}}^2)~.
\end{multline}
For brevity we omit the functional dependencies on the wavevector ${\bf k}$, time $t$ and $z$, as well as the species index $s$ wherever appropriate.

\begin{table}
	\setlength\extrarowheight{3pt}
	\caption{Material parameters of Al$_2$O$_3$ used for the static solution of the Boltzmann equation. The indices $i,e,h,*$ denote, respectively, ions and electrons in the plasma, and valence band holes and conduction band electrons in the wall.  }
	\label{tab:parameters}
	\begin{ruledtabular}
		\begin{tabular}{llll}
		$E_g(\text{eV})$                  &       6.24     &   	$m_{*,h}(m_e)$                    &       1      \\
		$E_t(\text{eV})$                  &   3.12  &       $m_i (m_e)$ 	&	1836      \\
		$N_t(\text{cm}^{-3})$	&	$10^{18}$& $k_B T_{i,*,h} (\text{eV})$         &       0.025    \\
		$\sigma_{*,h}(\text{cm}^2)$           &       $10^{-15}$ & $k_B T_{e} (\text{eV})$         &       2         \\
		$\varepsilon$                     &       3.27   & $\hbar\omega_0(\text{meV})$       &       48  \\
		$\varepsilon_\infty$              &       3.2     &&      \\
	\end{tabular}
\end{ruledtabular}
\end{table}

Before addressing the solution of Eq.~\eqref{eq:BE_1}, we comment on the unperturbed Boltzmann equation~\eqref{eq:BE_0} and its solution obtained by the method described in Ref.~\cite{RBF20}, focusing on the modifications necessary for a more realistic injection of the charge carriers at the interface.

In the solution of the unperturbed Boltzmann equation we include energy relaxation through polar optical phonons~\cite{Ridley99} as well as Shockley-Read-Hall recombination of electrons and holes~\cite{Hall51,SR52}. The expressions for $\Phi$ and $\gamma$ to which they lead are given in~\cite{RBF20}. 
For the calculations in this work we use Al$_2$O$_3$ as the prism material. The values for the band gap $E_g$, the  static and high energy dielectric constants $\varepsilon$ and $\varepsilon_\infty$, and the optical phonon energy $\hbar \omega_0$ are taken from the literature~\cite{Barker63,MC98}. 
As in Ref.~\cite{RBF20}, no particular impurity species was chosen for the Shockley-Read-Hall recombination process. The parameters for it are now however more realistic. In particular the trap density $N_t$ and capture cross sections $\sigma_{*,h}$ are now close to what one may expect in real Al$_2$O$_3$, while the trap state energy level $E_t$ is in the center of the band gap. 

The unperturbed Boltzmann equation~\eqref{eq:BE_0} is solved in a flat-band approximation, i.e. the band bending is neglected. 
This significantly simplifies the numerical treatment of Eq.~\eqref{eq:BE_0}, because it becomes an ordinary differential equation. 
In Ref.~\cite{RBF20} we included a self-consistent calculation of the potential profile, which we found however to be rather small. We neglect it thus for the purpose of this work. However, the injection processes are now modeled more realistically. For holes, a phenomenological, forward directed Gaussian injection is chosen, holes with total energy $E = k^2/(2m)$ and lateral kinetic energy $T_\textrm{lat} = (k_x^2+k_y^2)/(2m)$ are injected with the source function
\begin{equation}
S_h = n_0 \exp\left[-\frac{(E-I)^2 + T_\textrm{lat}^2}{\Gamma^2}\right]~,
\end{equation}
where the ionization energy $I = 13.6\,$eV (assuming a hydrogen plasma), and the energy width $\Gamma = 0.5\,$eV, mimicking the spread in energy for the holes generated by the neutralization of ions at the plasma-solid interface. The value of $\Gamma$ does not significantly influence the solution of the equations, and $n_0$ is chosen so that the flux is conserved.  Electrons are injected according to the quantum mechanical transmitivity of a potential step,
\begin{equation}
R_{\textrm{qm}} = \left(\frac{v_z^e-v_z^*}{v_z^e+v_z^*}\right)^2~,
\end{equation}
where $v_z^s$ is the $z$-component of the velocity of the electrons outside ($s=e$) or inside ($s=*$) the wall,
assuming a Maxwellian distribution for the electrons of the plasma. 
As in Ref.~\cite{RBF20}, the trap occupancy is not changed by the surplus charges. These approximations also apply for the first order solution discussed below.

\begin{figure}[t]
	\centering
	\includegraphics{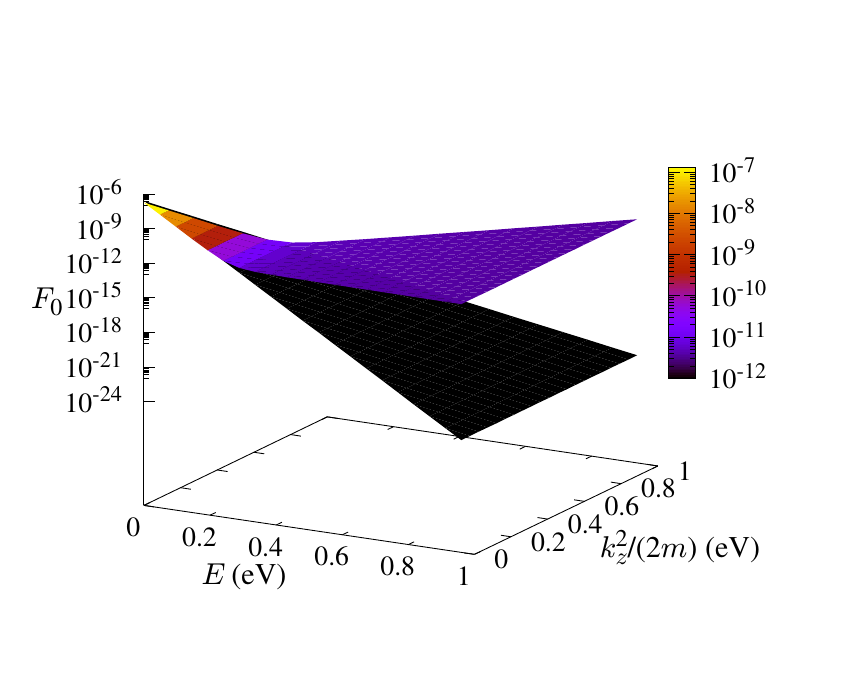}
	\caption{
		Representative solution of the static Boltzmann equation at the interface for low energies. Shown here is the distribution of holes at the interface ($z=0$), for $k_z<0$. As a reference in black is shown a Maxwellian distribution with equal density. Since the distribution is equivalent for small energies and quickly decreases, it should be sufficient to use a Maxwellian approximation when calculating the optical response. 
	}
	\label{fig:Boltzmann_solution}
\end{figure}

Shown in Fig.~\ref{fig:Boltzmann_solution} is a representative cut of the three dimensional unperturbed distribution function for holes at the interface. At other locations and for electrons the distributions have a similar shape. 
The most notable difference in the distribution function compared to the results presented in Ref.~\cite{RBF20} is that the distribution function at low energies is now almost a perfect sum of two Maxwellians, with $k_BT_*=0.025\,$eV, the wall temperature, and $k_BT_e=2\,$eV, the plasma-electron temperature. For higher energies features of the injection terms can be found, which are, however, some orders of magnitude smaller than the Maxwellian peak near $E=0$. Thus practically all weight of the function lies in the low temperature Maxwellian. Realistic injection energies are necessary for this effect to manifest, so that, for example, enough phonon collisions can occur during the relaxation process. In Ref. \cite{RBF20}, where the electrons and holes were injected at much lower energies, only about three phonons could be emitted before the bottom of the energy band was reached, now a hole, injected at $13.6\,\textrm{eV}$, can emit several hundred phonons.

While the band bending is not explicitly respected in the solution of the Boltzmann equation, the Poisson equation~\eqref{eq:PE} is still used to calculate the electric field and fix the strength of the plasma source through the matching condition of the field at the interface. This matching is equivalent to overall net charge neutrality across the electric double layer. Thus using the electric field without explicitly including the bending of the bands does not destroy this important property. 
The plasma side is treated collisionless, for more detail we refer to~\cite{RBF20} and~\cite{BF17}. By making the injection terms more realistic, some electrons on the plasma side are now reflected, which slightly modifies the equations for the plasma side compared to the calculations in the references above.
\begin{figure}[t]
	\centering
	\includegraphics{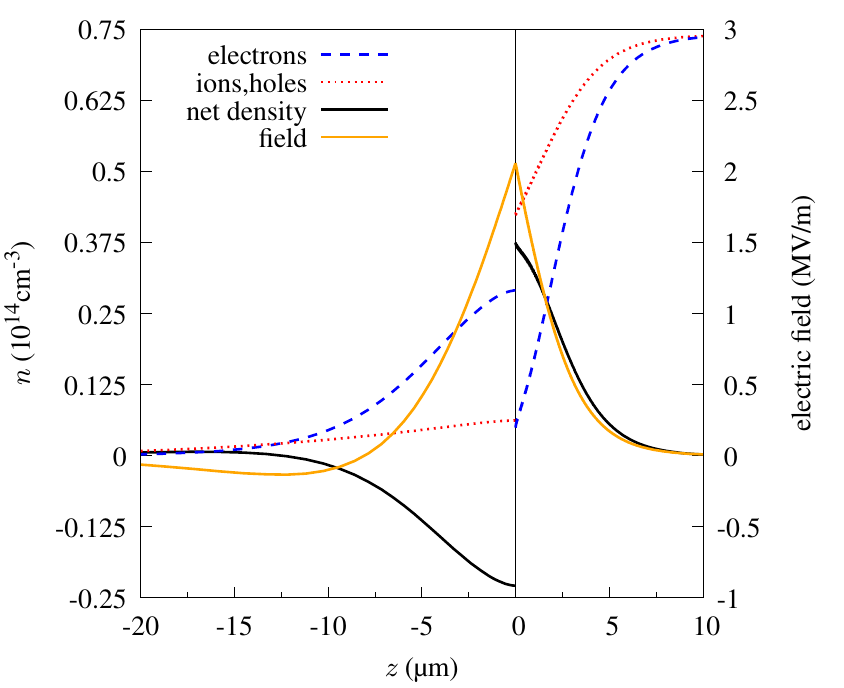}
	\caption{
		Static solution of the interface for Al$_2$O$_3$. On the solid side ($z<0$), the electric field is multiplied by the static dielectric function. As discussed in the main text the width of the plasma sheath is unrealistically small, but for the optical response this does not matter.
	}
	\label{fig:n&E}
\end{figure}

Figure \ref{fig:n&E} shows the density profiles and the electric field of the static solution for the Al$_2$O$_3$ interface.
Using the flat-band approximation and a collisionless plasma leads to a narrower plasma side of the electric double layer than the solid side, which is not what one would expect. 

On the plasma side, the collisionless model we employed leads to rather high plasma densities.
Thus the plasma sheath is narrow, because the length scale is given by the Debye length.
But since in our model the plasma only acts as a source for injecting electrons and holes into the solid, we found the simple model sufficient for our purposes.

Without band bending confining the charges to a narrow region at the interface, on the solid side the surplus charges spread rather deep in to the solid, causing a wide space charge region. 
Including the band bending would remedy this shortcoming.
However, the numerical resources required for solving Eqs.~\eqref{eq:BE_0} and \eqref{eq:PE} increase then dramatically, especially if the band bending is small compared to the injection energies. Since the width of the plasma sheath is less important to the optical response than the overall charge neutrality, which is obeyed, we decided not to take this additional numerical burden into account.
Using the results for the flat-band approximation we can, however, anticipate the effect band bending would have on the solution. 
Since the intrinsic electric potential is not very large, the collision integrals would still dominate the Boltzmann equation~\eqref{eq:BE_0}, thus favoring a Maxwellian shape of the distribution functions, like in the flat-band approximation. 
Hence, the densities would still be determined by the equilibrated Maxwellian background, rather than by the distributions of the injected charge carriers. For such a background, band bending would deplete holes and accumulate electrons close to the interface. 
Since the total net charge remains the same, due to the charge neutrality across the interface, the solid side of the electric double layer would then be significantly narrower. 
Even in the flat-band approximation, the density of holes is smaller than that of the electrons by about a factor five. Including band bending, this difference would be amplified. Thus, on the solid side essentially only electrons contribute to the space charge. 

Although the macroscopic solution obtained in the flat-band approximation with a collisionless plasma is thus not reliable, the properties of the microscopic distribution functions, mainly determined by the collision processes and the injection, should not be influenced by these approximations. 

We now turn to the solution of Eq.~\eqref{eq:BE_1} for the first order response of the distribution functions to the incident field.
Assuming harmonic dependencies on time and lateral directions, the Ansatz
\begin{equation}\label{eq:expFactor}
{\bf F}_1 = {\bf f}_1 \e^{ i \left( {\bf K}_F\cdot{\bf R} -\omega t\right)}~,
\end{equation}
where ${\bf K}_F = (k_{F}^x, k_{F}^y)^T$, ${\bf R} = (x,y)^T$ and $\omega$ is the frequency of the incident wave,
yields for Eq.~\eqref{eq:BE_1}
\begin{multline}\label{eq:BE_firstOrder}
	\left(-i\omega  + v_z\partial_z + {\bf V}\cdot {\bf K}_F + q \mathcal{E}_0 \partial_{k_z} +  \gamma_0  \right){\bf f}_1\\
	 = {\boldsymbol \varphi}_{1} - q\left[ {\boldsymbol \varepsilon}_1\cdot\hat{\bf z} + \delta(z-z')\right]\nabla_{{\bf k}}F_0~.
\end{multline}
Therein ${\boldsymbol{ \varphi}}_{1}$ and ${\boldsymbol \varepsilon}_1$ are ${\bf \Phi}_{1}$ and ${\boldsymbol{\mathcal{E}}}_1$, respectively, with the exponential factor of the type shown in Eq.~\eqref{eq:expFactor} split off, and ${\bf V} = (v_x,v_y)^T$. 

The flux is calculated from the distributions $F_s$ as
\begin{equation}
\label{eq:j}
{\bf j}(z) = \int\frac{\rmd^3k}{(2\pi)^3} \left[F_h(z,{\bf k}){\bf v}_h({\bf k})-F_*(z,{\bf k}){\bf v}_*({\bf k})\right]~.
\end{equation}
Only the first order carriers are responsible for the response to the incident field, thus the flux of the unperturbed solution does not appear in Eq.~\eqref{eq:j}. Eqs.~\eqref{eq:j=simgaE},~\eqref{eq:F_expansion} and~\eqref{eq:j} enable one to identify 
\begin{equation}\label{eq:sigma}
\underline{\underline{\sigma}}(z,z') = q\int\frac{\rmd^3k}{(2\pi)^3} {\bf v}({\bf k}) \otimes {\bf f}_1(z,z',{\bf k})~,
\end{equation}
where $\otimes$ denotes the outer product. Not to overload the expressions, we give in the following the equations only for one species and note that the contributions of holes and electrons are added for the final result.

Analyzing Eq.~\eqref{eq:BE_firstOrder} for parameters present in plasma-solid interfaces, the dominant terms are $\omega$ on the left and $\delta(z-z')$ on the right side.
Since $\abs{{\bf K}_f}\propto \omega/c$, and $v$ in a.u. is, for the relevant energies, typically smaller than one, the term ${\bf V}\cdot {\bf K}_F$ can be neglected for a first approximate solution. From the results for the unperturbed Boltzmann equation, the length scale on the solid side is about the same order of magnitude as the wavelength of the infrared light, making the drift term $v_z\partial_z$ in Eq.~\eqref{eq:BE_1} the same order of magnitude as the ${\bf V}\cdot {\bf K}_f$ term.
The electric field is neglected in the flat-band approximation, and the collision terms, $\gamma_0 {\bf f}_1$ and ${\bf \phi}_1$, are for the collision processes considered here also small compared to $\omega{\bf f}_1$, by about two and four orders of magnitude for phonon collisions and trap recombination, respectively. 
Using the method of dominant balance,
\begin{equation}
{\bf f}_1 = -i\frac{q}{\omega}\nabla_{{\bf k}}F_0\delta(z-z')~
\end{equation}
is thus a reasonable approximation for the solution of Eq.~\eqref{eq:BE_firstOrder}. Hence the response is in leading order local in $z$. 
Moreover, if $F_0$ is assumed to be Maxwellian, as in fact it nearly is, see Fig.~\ref{fig:Boltzmann_solution}, the resulting conductivity becomes
\begin{equation}\label{eq:Drude}
\sigma(z) = i\frac{q^2n(z)}{m\omega}~
\end{equation}
for both $\sigma_{xx}$ and $\sigma_{zz}$. The density $n(z)=n_*(z)+n_h(z)$, since both electrons and holes add to the conductivity. 
Having found that the distributions are effectively Maxwellian, and that the density of holes can be neglected, it is sufficient to use only the electron density in Eq.~\eqref{eq:Drude}.
We have thus shown that it is justified to use the local Drude model to calculate the transmitivity of the MIRE, as we did in our exploratory calculation~\cite{BRF20}.

\begin{figure}[t]
	\centering
	\includegraphics{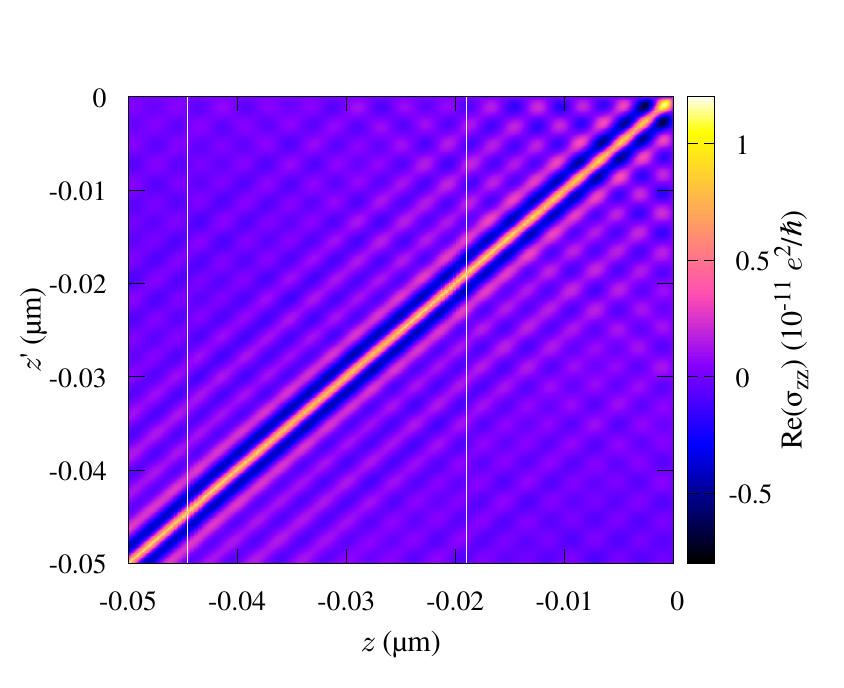}
	\caption{
		Real part of the nonlocal conductivity $\sigma_{zz}(z,z')$ for Al$_2$O$_3$. Oscillatory behavior is clearly recognizable perpendicular to the diagonal. There are also oscillations in $z+z'$ direction, which are caused by the term that comes from the reflection at the interface, i.e. the second term on the right of Eq.~\eqref{eq:f1<}. The behavior of the imaginary part and $\sigma_{xx}(z,z')$ is similar.
		Note the different scales from Fig.~\ref{fig:n&E}.
	}
	\label{fig:2D_sigma}
\end{figure}
To verify the validity of the local approximation, we calculated the first correction to the local solution by including the drift term $v_z\partial_z$, the term ${\bf V}\cdot {\bf K}_F$ and the term $\gamma_0^\textrm{rec}$ from the recombination processes, making the solution nonlocal, while keeping the resulting equation
\begin{multline}\label{eq:BE_nonlocal_first_order}
\left(-i\omega  + v_z\partial_z + {\bf V}\cdot {\bf K}_F + \gamma_0 ^\textrm{rec} \right){\bf f}_1\\
= - q \delta(z-z')\nabla_{{\bf k}}F_0~
\end{multline}
analytically solvable.

Due to the different boundary conditions, it is convenient to solve Eq.~\eqref{eq:BE_firstOrder} separately for motion towards and away from the interface, which is indicated by an upper index $>$ and $<$, respectively. 
Using ${\bf f}_1^>(z_1,z') =0$, where $z_1$ is a point far away from the interface, where the behavior is effectively that of the bulk, one finds
\begin{equation}
{\bf f}_1^>  (z,z')= -\frac q {v_z} \nabla_{{\bf k}}F_0^>(z')  I(z,z')~,
\end{equation}
while the specular reflection boundary condition at the interface, ${\bf f}_1^<(0, z') = {\bf f}_1^>(0,z')$, yields
\begin{multline}
\label{eq:f1<}
{\bf f}_1^<  (z,z')= \frac q {v_z} \nabla_{{\bf k}}F_0^<(z') I(z',z)\\
-\frac q {v_z} \nabla_{{\bf k}}F_0^>(z') I(0,z)I(0,z')
\end{multline}
with
\begin{equation}
I(z,z') = \exp\left( -\int_{z'}^z\rmd z''\frac{\tilde{\gamma}}{v_z} \right)\Theta(z-z')
\end{equation}
and
\begin{equation}
\tilde \gamma = -i\omega + {\bf V}\cdot {\bf K}_F + \gamma_0^\textrm{rec}~.
\end{equation}
Here, $v_z$ is the unsigned velocity in $z$ direction.

Figure \ref{fig:2D_sigma} shows part of the nonlocal conductivity $\underline{\underline{\sigma}}$ calculated with Eq.~\eqref{eq:sigma} using this solution. It is strongly peaked around $z =z'$, with oscillations of small amplitude on a very short lenghtscale away from it.
The phonon collisions that were omitted in Eq. \eqref{eq:BE_nonlocal_first_order}, which only manipulate the momentum variables, would not introduce any more nonlocality. 
Since in the integrations over the nonlocal conductivity in Eqs. \eqref{eq:epsxx(z)} and \eqref{eq:epszz(z)} the oscillations are effectively negligible compared to the peak on the diagonal, this confirms that for the parameters at the plasma-solid interface a local approximation for the conductivity is sufficient for the calculation of the surface response functions.

\section{Results}\label{sec:results}

Having demonstrated that the local Drude term is sufficient for calculating the optical response, we present in this section results for the transmitivity of the MIRE using this approximation.
Inserting it into Eq.~\eqref{eq:d_para} immediately shows that the surface response function $d_\parallel$ only depends on the integrated net surface charge
\begin{equation}
n_S = \int_{-\infty}^0 \left[n_*(z)-n_h(z)\right]\rmd z \approx \int_{-\infty}^0 n_*(z)\rmd z
\end{equation}
according to
\begin{align}\label{eq:d_para_final}
	d_\parallel = \frac{1}{1 - \varepsilon}\frac{4\pi q^2 n_S}{m\omega^2}~.
\end{align}
In calculating $d_\perp$, which includes the inverse of the nonlocal dielectric function, the local approximation also leads to a significant simplification. To find the inverse one no longer needs matrix inversion in Eq.~\eqref{eq:inv_eps}. Instead, the local inverse applies.  
For $\frac{4\pi q^2n(z)}{m\omega^2}\ll \varepsilon$, which is easily fulfilled, Eq.~\eqref{eq:d_perp} becomes
\begin{equation}\label{eq:d_perp_final}
d_\perp =  \frac{1}{\frac{1}{\varepsilon} - 1}\frac{4\pi q^2n_S}{m \omega^2 \varepsilon^2} + \mathcal{O}\left[\left(\frac{4\pi q^2n(z)}{m\omega^2\varepsilon}\right)^2/\varepsilon\right]~.
\end{equation}
The specific spatial charge distribution only plays a minor role in the calculation of the surface response functions, and can be neglected in leading order. Thus, the infrared response of the electric double layer can effectively be described by the integrated surface charge as a parameter. While this is perhaps the first approximation one would make, as we indeed did in~\cite{BRF20}, the preceding calculations show the validity of such a model.

So far we have not included the plasma side in the calculation of the surface response functions. In the plasma sheath the positive space charge is caused by a surplus of ions close to the interface, and since Eqs.~\eqref{eq:d_para_final} and~\eqref{eq:d_perp_final} show that the surface response functions are antiproportional to the mass of the charge carriers, the contributions of the ions are negligible compared to the electrons in the wall forming the negative part of the double layer.

\begin{figure}[t]
	\centering
	\includegraphics{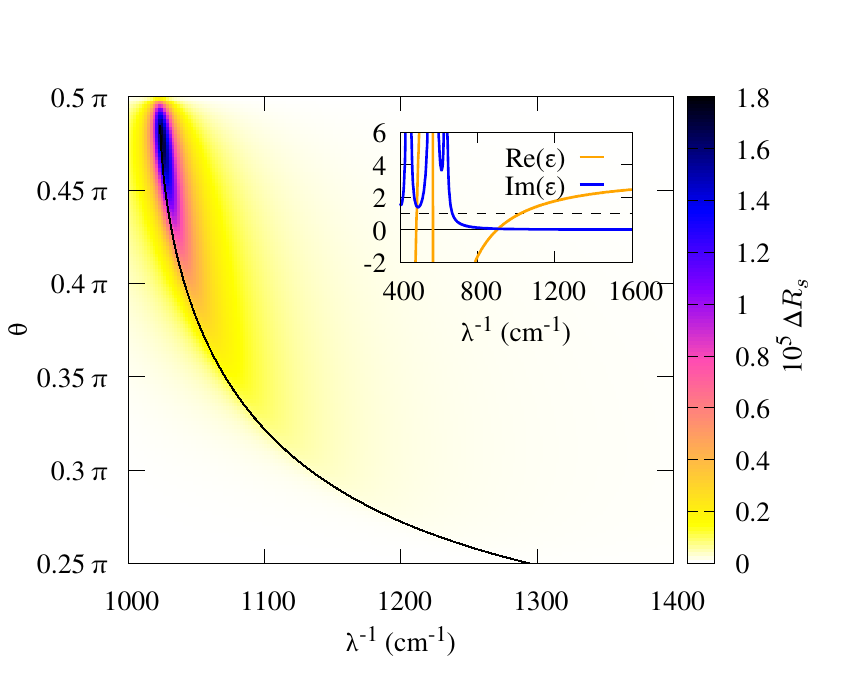}
	\caption{
		The change in the reflectivity $\Delta R_s$ at the Al$_2$O$_3$ interface. 
		The black line indicates the critical angle for total reflection.
		Also shown is the dielectric function in the infrared spectral range.
	}
	\label{fig:2D_R}
\end{figure}

\begin{figure}[t]
	\centering
	\includegraphics{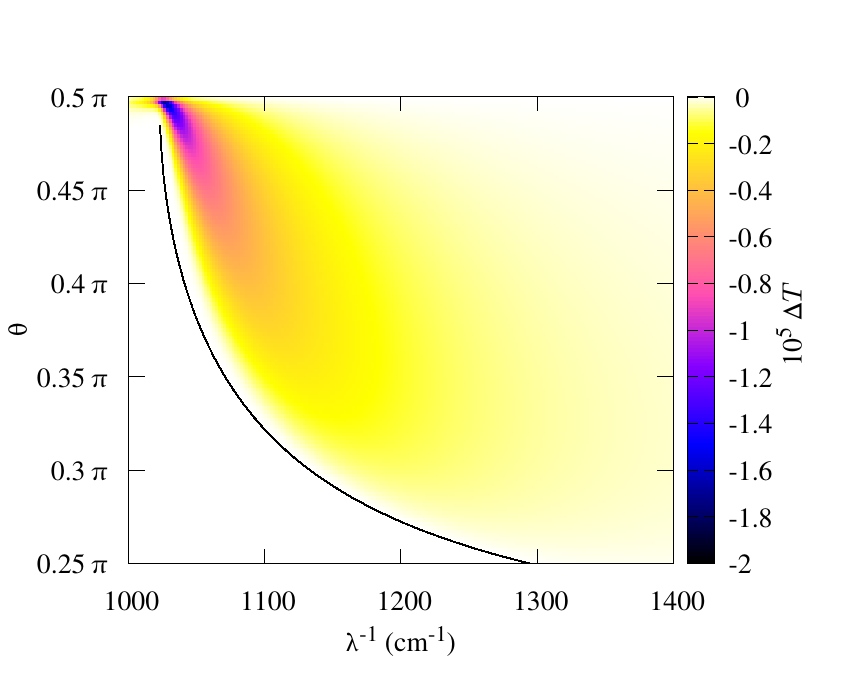}
	\caption{
		Change in the transmitivity $\Delta T = T-T_0$ of the Al$_2$O$_3$ MIRE for $N = 13$. $T_0$ denotes the transmitivity without the influence of the plasma, calculated from Eq.~\eqref{eq:T} with $C = 0$.
		The black line indicates the critical angle for total reflection.
	}
	\label{fig:2D_transmitivity}
\end{figure}

\begin{figure}[t]
\centering
\includegraphics{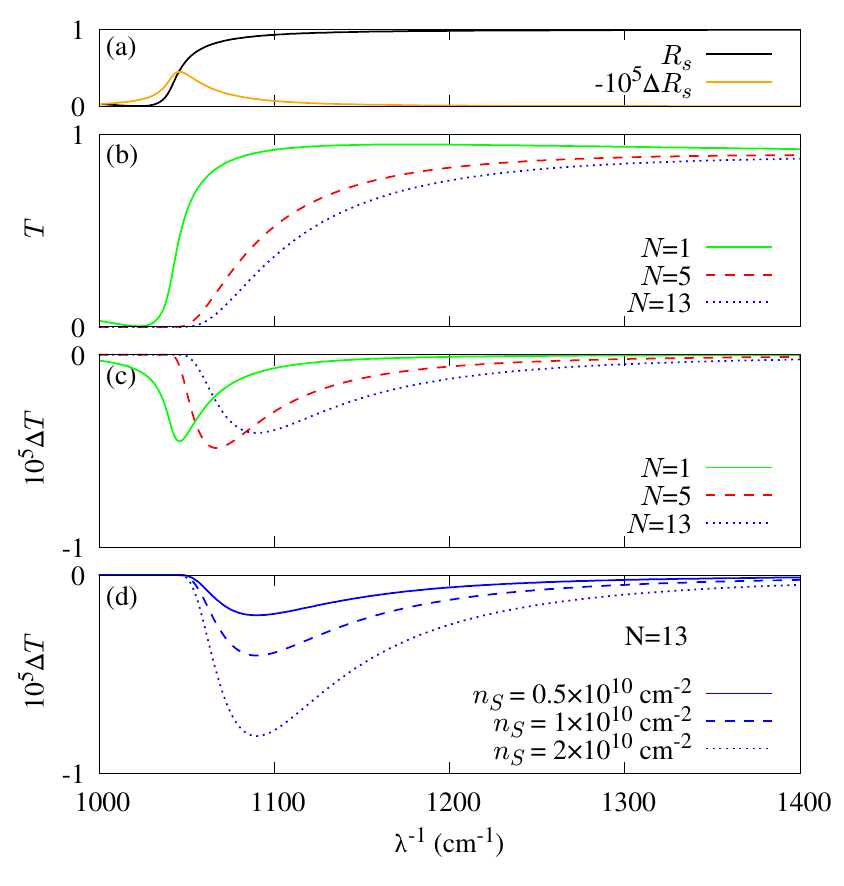}
\caption{
	For $\theta = 0.4\,\pi$, (a) surface reflectivity of the plasma-solid interface and change $\Delta R_s$ caused by the surplus charges, (b) transmitivity of the MIRE for different numbers of reflections $N$, (c) change in the transmitivity through the surplus charges for different $N$ with $n_S = 10^{10}\,\textrm{cm}^{-2}$, and
	(d) change in the transmitivity for different surface charges $n_S$.
}
\label{fig:1D_transmitivity}
\end{figure}
Figure \ref{fig:2D_R} shows the change of the reflection coefficient $\Delta R_s = \abs{r}^2-\abs{r_0}^2$ as a function of the angle of incidence and the inverse wavelength for a surface charge density $n_S = 10^{10}\textrm{cm}^{-2}$.
Notably, the largest changes occur in a narrow region close to the critical angle of total reflection, indicated by the black line. The inset of Fig.~\ref{fig:2D_R} shows the bulk dielectric function of the prism material Al$_2$O$_3$. Only for energies where the dielectric function is larger than one, indicated by the dashed line, can total internal reflection occur, so the limit for total reflection is at about $1050\,$cm$^{-1}$. The bulk dielectric function is modified through the Drude term~\eqref{eq:Drude}, so expectedly the largest change in the reflectivity occurs near the critical angle, which, by the Drude term, gets shifted. 

Figure \ref{fig:2D_transmitivity} shows the resulting change in the transmitivity of the MIRE.
While the magnitude of the change is not considerably larger, as one may have expected, the feature is spread over a much wider spectral range, and is thus easier to measure.

Figure \ref{fig:1D_transmitivity} illustrates the effect of the MIRE on the measured signal. Figure \ref{fig:1D_transmitivity}(a) shows the reflectivity of the plasma-solid interface as well as the change caused by a surface charge $n_S = 10^{10}\,\textrm{cm}^2$.
In Fig.~\ref{fig:1D_transmitivity}(b) the transmitivity of the MIRE is shown for different numbers of internal reflections. For $N=1$ the transmitivity closely resembles the reflectivity of the plasma-solid interface, since $R_0\approx 1$. 
With raising numbers of reflections $N$ the deviations of $R_s$ from 1 at high wavelengths are amplified, hence the transmitivity at higher energies becomes smaller and the transition from $T\approx0$ to $T\approx 1$ is broadened.
The change of the transmitivity that is induced by the surplus charges, displayed in Fig.~\ref{fig:1D_transmitivity}(c), while not significantly gaining or losing amplitude, is consequently also spread over a wider spectral range.
This change, as can be seen from Fig.~\ref{fig:1D_transmitivity}(d), depends approximately linearly on the integrated surface charge, allowing thus the measurement of the charge in a straightforward way. 

Since the optical response can be calculated using only the integrated surface charge, the influence of additional surface effects, such as an adlayer of adsorbed molecules, can be integrated into the model.
Without the previous conclusions, such effects would need to be  included into the boundary conditions of the Boltzmann equations, as well as, if electrically charged, into the Poisson equation. 
For a dilute adsorbate the dielectric function can in the simplest form be modeled by a single vibrational resonance $\omega_T$ with damping $\gamma$ and a strength $\omega_P$, giving rise to the dielectric function~\cite{IbachMills82}
\begin{equation}
	\varepsilon_\textrm{ad} (\omega) = 1 + \frac{\omega_P^2}{\omega_T^2 - \omega^2 - i \gamma \omega}~.
\end{equation}
For an adsorbate layer of thickness $d_\textrm{ad}$ at the wall facing the plasma, the contributions to the surface response functions are
\begin{equation}
	d_\parallel^\textrm{ad} = \frac{\varepsilon_\textrm{ad}(\omega)-1}{\varepsilon(\omega)-1}d_\textrm{ad}
\end{equation}
and 
\begin{equation}
	d_\perp^\textrm{ad} = \frac{\varepsilon_\textrm{ad}^{-1}(\omega)-1}{\varepsilon^{-1}(\omega)-1}d_\textrm{ad}~,
\end{equation}
which are added to Eqs.~\eqref{eq:d_para_final} and~\eqref{eq:d_perp_final}, respectively. 
When the resonance frequencies of the adsorbed molecules are far outside the relevant spectral range for the charge measurement, these contributions do not influence the resulting signal significantly. 

\begin{figure}[t]
	\centering
	\includegraphics{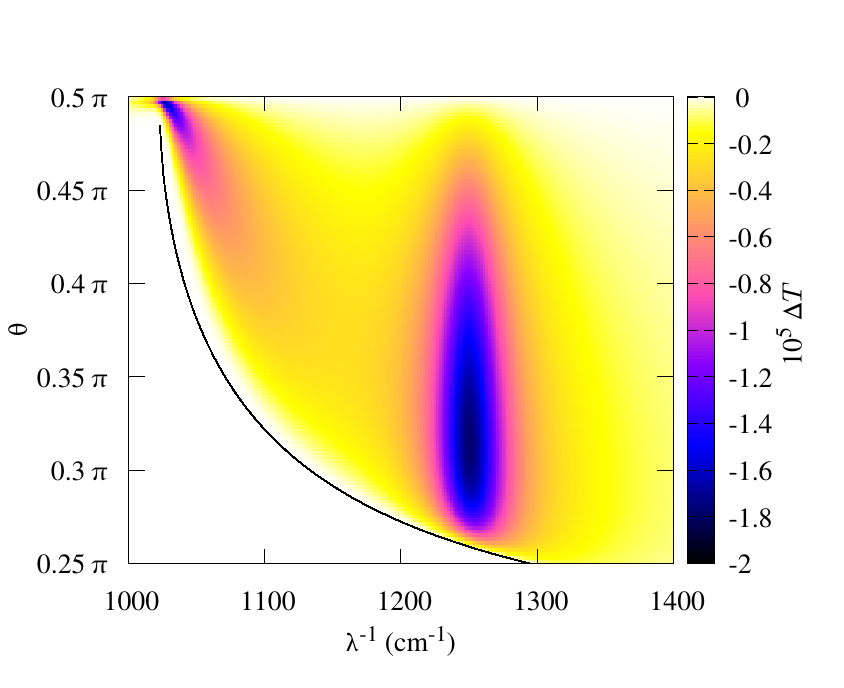}
	\caption{
		Same as Fig. \ref{fig:2D_transmitivity}, but including an adlayer as described in the main text.
	}
	\label{fig:2D_transmitivity_adlayer}
\end{figure}

To illustrate the effect of the adlayer and as a proof of principle, Fig.~\ref{fig:2D_transmitivity_adlayer} shows the transmitivity of the MIRE with an adlayer  for $d_\textrm{ad} = 0.3\,$nm, $\omega_T = 1250\,\textrm{cm}^{-1}$,  $\omega_P = 10\,\textrm{cm}^{-1}$, and $\gamma = 50\,\textrm{cm}^{-1}$.
The effect of the adlayer is confined to the region around $\omega_T$, where the dielectric function deviates from 1. The spectral range used for the charge measurement, around $\omega = 1050\,\textrm{cm}^{-1}$, is thus not affected by the presence of the adlayer.
The parameters do not represent a specific adlayer, but are chosen so that the effect can clearly be identified in Fig.~\ref{fig:2D_transmitivity_adlayer}. The resonance frequency of an adsorbed CO molecule, for example, is $\omega_T= 2149\,\textrm{cm}^{-1}$, with $\gamma = 10\,\textrm{cm}^{-1}$~\cite{YWN20}. Such an adlayer would not cause any noticeable change for the energies shown.

\section{Conclusion}\label{sec:conclusion}

In this paper we have shown that the dielectric response of the electric double layer at a plasma-facing wall is Drude-like. Our analysis of the distribution function shows a Maxwellian behavior for the unperturbed charge carriers, which allows a local approximation for the conductivity.
When using a local Drude conductivity in the surface response functions, the spatial charge distribution plays only a minor role, $d_\parallel$ does not depend on the spatial distribution of the charges at all, and $d_\perp$ only depends on it in higher orders.
Thus, the optical response of the plasma-facing solid is in leading order a function of only the integrated surface charge and knowledge of the spatial distribution is not necessary to determine the absorption, nor can the proposed measurement provide information about it. Nonetheless the present work confirms that the surface charge, which itself is of much interest, can be measured in the proposed configuration.

Compared to our previous proposal for infrared spectroscopy of the wall charge~\cite{RBB18}, where avoided crossing of the Berreman mode and the resonance of a surface plasmon polariton was utilized to measure the surface charge through the shift of the reflectivity minimum, the configuration in this work is much simpler.
In the stack approach of Ref.~\cite{RBB18} a metallic layer on the prism is necessary for the surface plasmon resonance, and the plasma-facing dielectric layer needs to be thin, in order to host the Berreman mode. In order to prevent spill over of electrons from the dielectric to the metal layer, we suggested an additional, electro-negative layer between the metal and the plasma-facing dielectric. This also justified a model where the surplus charges are homogeneously distributed within the dielectric layer, because the calculation could at the time not account for inhomogeneous charge distributions. 

While the change in the transmitivity is relatively small, with about $10^{-5}$ for $n_S = 2\times 10^{10}\,\textrm{cm}^{-2}$, it should be measurable. MIREs have already been used for changes of about $10^{-3}$ decades ago~\cite{NM97, LKA93}, so we expect modern instruments to provide the necessary sensitivity.
Another experimental configuration that is very sensitive to small changes in surface reflectivities is the Cavity-Ring-Down method~\cite{KD88, MMB94,BPM00}, where a pulsed signal is reflected numerous times at the interface within a cavity and information about the reflectance is found from decay time measurements of the signal exiting the cavity. 
In principle it could therefore also be employed to measure the accumulated charges at the plasma-solid interface. 
However, the unperturbed reflectivity $\abs{r_0}^2$ must then be very close to one, so that the deviations are of the same order of magnitude as the change induced by the surplus charges.
This may not be the case for all materials of interest. For instance, Al$_2$O$_3$ could not be used in a cavity ring down experiment for this purpose. The MIRE approach does not impose such a strong restriction on the choice of the wall material.

We have also shown that surface impurities, such as an adlayer of adsorbed molecules, do not negatively affect the charge measurement, provided the resonances are in a different spectral range. It should be noted that the MIRE method could, if applied at the corresponding energies, also be used for in-operando diagnostics of such adlayers. 

\section*{Acknowledgments}
Support by the Deutsche Forschungsgemeinschaft through project BR-1994/3-1 is greatly acknowledged.

\bibliography{ref}

\end{document}